\documentclass[preprint,showpacs,showkeys,preprintnumbers,amsmath,amssymb]{revtex4}

\usepackage{graphicx}
\usepackage{dcolumn}
\usepackage{bm}

\begin{document}

\title{Interface relaxation and electrostatic charge depletion in the oxide heterostructure LaAlO$_3$/SrTiO$_3$}

\author{U.~Schwingenschl\"ogl$^{1,2}$ and C.~Schuster$^2$}
\affiliation{
$^{1}$ICCMP, Universidade de Bras\'ilia,
    70904-970 Bras\'ilia, DF, Brazil \\
$^{2}$Institut f\"ur Physik, Universit\"at Augsburg, 86135 Augsburg, Germany}

\date{\today}

\begin{abstract}
Performing an analysis within density functional theory, we develop insight into
the structural and electronic properties of the oxide heterostructure
LaAlO$_3$/SrTiO$_3$. Electrostatic surface effects are decomposed from the internal
lattice distortion in order to clarify their interplay. We first study the interface
relaxation by a multi-layer system without surface, and the surface effects, separately,
by a substrate-film system. While elongation of the TiO$_6$ octahedra at the
interface enhances the metallicity, reduction of the film thickness has the
opposite effect due to a growing charge depletion. The interplay of these two effects,
as reflected by the full lattice relaxation in the substrate-film system, however,
strongly depends on the film thickness. An inversion of the TiO$_6$ distortion
pattern for films thinner than four LaAlO$_3$ layers results in an insulating state.
\end{abstract}


\maketitle

Perovskite heterostructures from transition metal oxides
\cite{kroemer01,imada98} have attracted recent interest due to the
discovery of metallic inter-layers in an otherwise semiconducting
system. For example, a metallic contact between two common band
insulators, with large band gaps of 5.6 eV and 3.2 eV, is realized
at the LaAlO$_3$/SrTiO$_3$ heterointerface \cite{ohtomo04}. From a
structural point of view, it consists of (SrO)$^0$, (TiO$_2$)$^0$,
(LaO)$^+$, and (AlO$_2$)$^-$ layers. Because of different formal
valences of the metal Ti$^{4+}$ and La$^{3+}$ ions, the
(TiO$_2$)$^0$/(LaO)$^+$ interface is electron-doped and a
two-dimensional electron gas is formed. Nevertheless, it has been shown
experimentally that the LaAlO$_3$ surface layer must reach a critical
thickness of 4 unit cells for the interface to become conducting
\cite{thiel06}. The phenomenon of metallicity has to be attributed
to the electronic relaxation, which usually is accompanied by a
fundamental lattice relaxation. Local deviations from the bulk
crystal structure as well as a transfer of charge across the
interface, induced by different electro-chemical potentials in the
component materials, hence plays a key role for the formation of
conduction states \cite{okamoto04,herranz07}.

Pulsed laser deposition techniques and molecular beam epitaxy nowadays, in
principle, make it possible to grow layered structures with a precision of
a single unit cell and to create atomically sharp interfaces. Close to contacts
between perovskite oxides the local structural and electronic properties 
often differ strikingly from the bulk materials, similar to the reconstruction
and formation of specific states at surfaces. While the importance of relaxation
for understanding heterointerfaces has been stressed in various studies
\cite{hamann06,willmott07,pentcheva08,zhong08}, very little is known about a
possible interplay with a nearby surface, which, depending on the experimental
setup, can be of high relevance.

In our present paper we aim at providing these informations for the
LaAlO$_3$/SrTiO$_3$ heterointerface. We organise our considerations as follows:
After settling structural details of the systems under investigation
and specifying the computational procedure, we first discuss the
interface effects on the electronic properties of multi-layer
samples. Next we study surface effects on the interface-relaxed
system and discuss a band bending picture for explaining
charge depletion under variation of the film thickness. As final
step we further relax the full system (interface and surface) to
shed light on the crucial interplay between the surface electrostatics
and the internal lattice distortion.

In our calculations, structural relaxation at both the LaAlO$_3$/SrTiO$_3$
interface and in the LaAlO$_3$ film is taken into account by means
of a minimisation of the atomic forces. The electronic structures and
relaxation patterns are obtained by the Wien2k program within the generalised
gradient approximation \cite{wien2k}. This implementation of density functional
theory makes use of a mixed linear augmented-plane-wave and augmented-plane-wave plus
local-orbitals basis. It has shown a great capability for studying interfaces
\cite{interf2,interf3,interf} and surfaces \cite{surf2,surf3,surf}. As parametrisation
of the exchange-correlation potential we employ the Perdew-Burke-Ernzerhof scheme.
This approximation results in the well-known underestimation of the band gaps of
semiconductors and insulators, amounting to only 3.4 eV and 2.1 eV for bulk
LaAlO$_3$ and SrTiO$_3$, respectively. Our basis set consists of the valence states
La $6s$, $6p$, $5d$, Sr $5s$, $5p$, Ti $4s$, $4p$, $3d$, Al $3s$, $3p$ and O $3s$,
$2p$, and the semi-core states La $5s$, $5p$, Sr $4s$, $4p$, Ti $3s$, $3p$, Al $2p$,
and O $2s$.

Our discussion relies on two structural setups: (1) an artificial {\it multi-layer}
configuration without surface, used for separating the surface from the interface
effects, and (2) the {\it substrate-film} system, which finally includes a surface.
The multi-layer supercell consists of two LaAlO$_3$ unit cells, which are sandwiched
between 9 SrTiO$_3$ unit cells on each side. The stacking in c-direction hence follows
a \ldots-- 18 SrTiO$_3$ -- 2 LaAlO$_3$ --\ldots\ scheme. It has been checked
that the SrTiO$_3$ slab is thick enough to be considered bulk-like.
Furthermore, we assume that no superstructure is formed in the $ab$-plane,
which is likely due to similar lattice constants. In summary, the supercell
contains 42 inequivalent sites. The lattice constant is set to 3.905\,\AA\ (value of
bulk SrTiO$_3$; both within the $ab$-plane and in the $c$-direction) and not optimized
systematically. Variation of this value between the lattice constants of the
component materials, however, does not change the following findings. We deal with an
n-type contact with LaO/TiO$_2$ stacking at the interface. Our {\bf k}-space grid
(12,12,1) contains 21 points within the irreducible wedge of the Brillouin zone.
To represent the charge density we use 27000 plane waves, where
the cutoff is RK$_{\rm max}=5.5$.

The substrate-film system is modelled by means of an inversion symmetric
SrTiO$_3$ layer which is 5 unit cells thick and covered by layers of two to four
LaAlO$_3$ unit cells on its top and bottom. Again, LaO/TiO$_2$ stacking at the interface
results in an n-type contact. The LaAlO$_3$ film is terminated by an AlO$_2$ plane.
Using a fixed length of the supercell, the vacuum region has a thickness from
12\,\AA\ to 27\,\AA. Here, the {\bf k}-space grid (11,11,1) ensures 21 points
within the irreducible wedge of the Brillouin zone. Finally, the charge density is
represented by 14000 plane waves with cutoff RK$_{\rm max}=7.0$

The lattice distortion at the contact between LaAlO$_3$ and SrTiO$_3$ is
characterised by displacements of the anions and cations in opposite directions,
because of Jahn-Teller distorted (elongated) TiO$_6$ octahedra \cite{vonk07}. In the literature,
results have been reported for both the LaAlO$_3$/SrTiO$_3$ \cite{park06,gemming06,albina07,maurice08}
and the LaTiO$_3$/SrTiO$_3$ interface \cite{hamann06,okamoto06}. Within the SrTiO$_3$ substrate,
the lattice distortions decay exponentially \cite{ussub}. Electron doping of the
Ti atoms, in combination with the Jahn-Teller effect, leads to the appearance of
metallicity in band structure calculations relying on the local density approximation
(LDA). Considering the local electron-electron interaction more accurately by means
of the LDA+U scheme, Pentcheva and Pickett \cite{interf3} have succeeded in
obtaining a ferromagnetic spin order tracing back to occupied $d_{xy}$ states
in a Ti$^{3+}$ checkerboard. Moreover, it has been demonstrated that the conducting
LaAlO$_3$/SrTiO$_3$ interface states are restricted to a region of only two SrTiO$_3$
unit cells perpendicular to the contact \cite{ussub,reyren07,basletic08}.

\begin{figure}[ht!]
\flushleft{\large (a)}\\
\includegraphics[width=0.37\textwidth,clip]{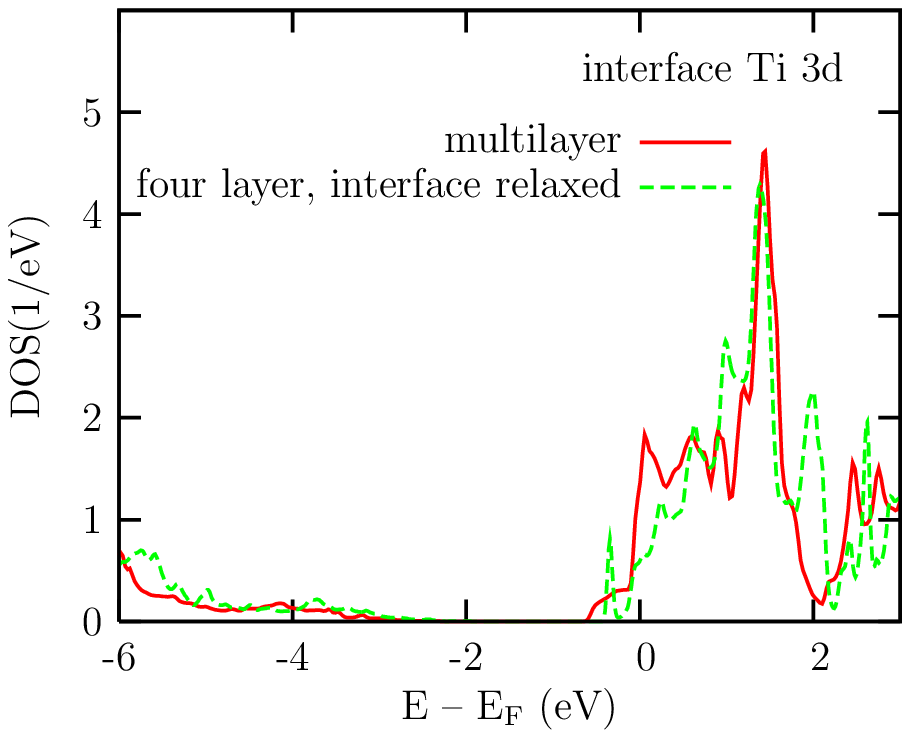}
\vspace{-0.5cm}
\flushleft{\large (b)}\\
\includegraphics[width=0.37\textwidth,clip]{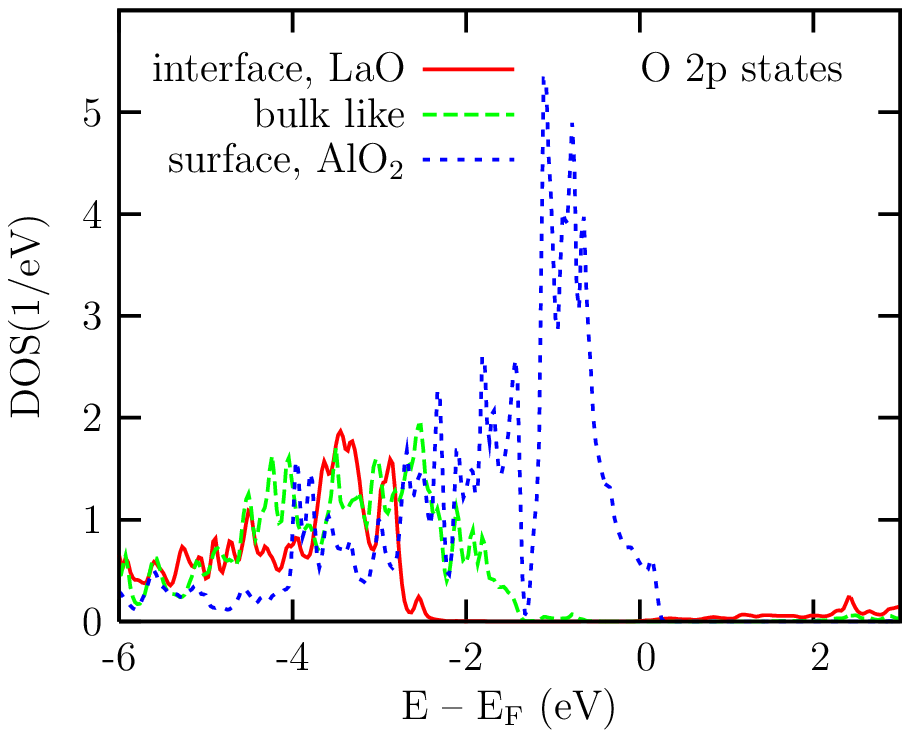}
\vspace{-0.5cm}
\flushleft{\large (c)}\\
\includegraphics[width=0.37\textwidth,clip]{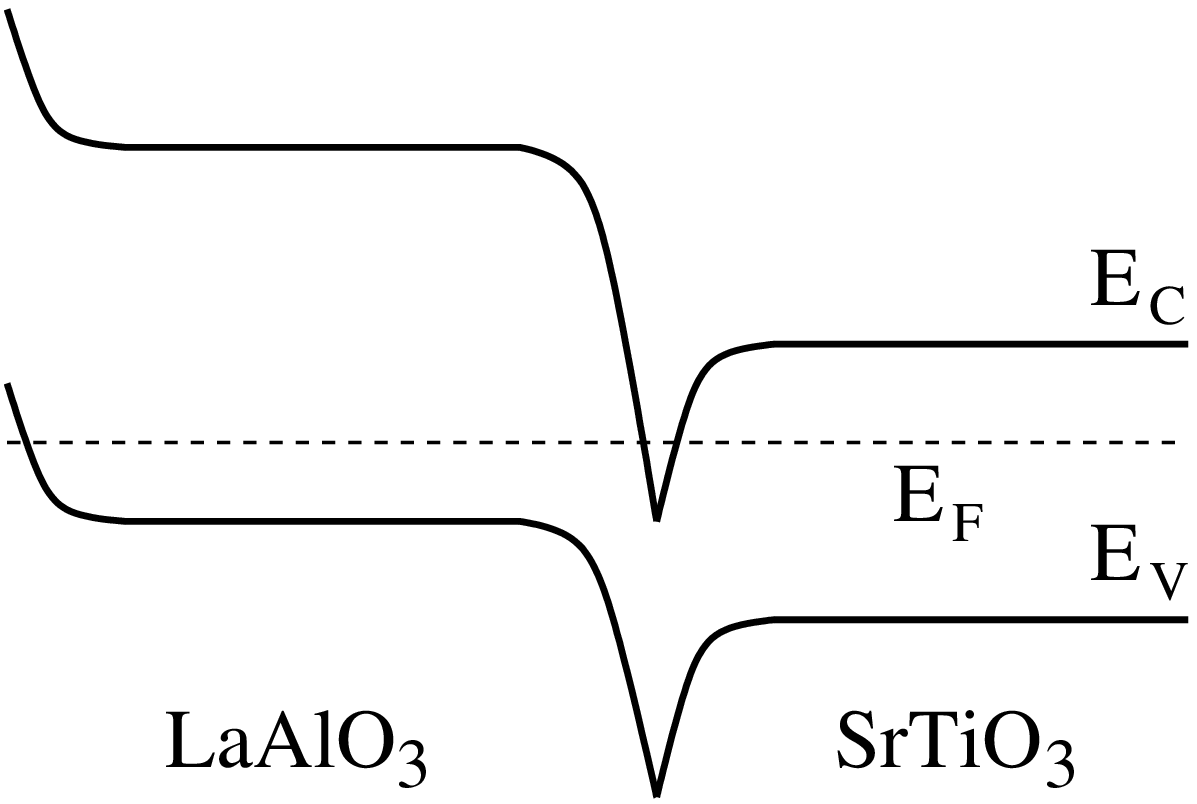}
\caption{(a) Interface Ti $3d$ DOS for the multilayer system and the four layer
system, accounting only for the interface structure relaxation. A down bending of the
Ti $3d$ bands below the Fermi level is evident. The computed band gap reproduces
the bulk SrTiO$_3$ data. 
(b) Comparison of the O $2p$ states at the interface, the surface, and in a
bulk-like LaO layer of the four layer system. 
(c) Schematic illustration of the band bending in the heterostructure.}
\label{fig_DOS_bend}
\end{figure}

Starting from the structurally relaxed heterointerface, we now investigate
the influence of the surface by taking into account relaxation at the
vacuum interface. We find that the insulating state of an epitaxial
LaAlO$_3$/SrTiO$_3$ heterointerface near a surface is fully explained
by first-principles electronic structure calculations \cite{us08}. On variation of the
LaAlO$_3$ layer thickness on the SrTiO$_3$ substrate, the interface
conduction states are subject to an almost rigid band shift tracing back to a modified
Fermi energy. Since surface electronic states leak out into the vacuum
(the charge carrier density is reduced in the LaAlO$_3$ surface layer),
less charge is available for the interface doping. To interpret these facts
in terms of a band bending scheme, similar to semiconductor heterojunctions,
we have to analyze the projected density of the states near the interface and
surface. At semiconductor heterojunctions the band bending
yields a potential wall and the interface becomes insulating, as the
electronic bands do not cross the Fermi level in the contact region. Since
a pn-junction likewise has a depletion zone, comparability to typical
semiconductor devices, of course, is limited.

As mentioned above, the Ti $3d$ bands at the interface are shifted to lower
energies due to charge transfer off the LaO layer, and conduction states are formed. 
A similar behaviour is observed for the substrate-film system with four LaAlO$_3$
layers, for which the partial Ti DOS is shown in Fig \ref{fig_DOS_bend}a.
For a heterojunction one expects that the O $2p$ states in the LaAlO$_3$
layer on the opposite side of the contact likewise reveal a down bending. This
expectation is confirmed by Fig.\ \ref{fig_DOS_bend}b, full line. In addition,
a distinct up bending of the O bands occurs at the surface, see the dotted line
in Fig \ref{fig_DOS_bend}b. Combining these two effects we arrive at the schematic
band bending scheme for our heterointerface in Fig.\ \ref{fig_DOS_bend}c,
which should give an overview of the scenario discussed in the following.

\begin{figure}[ht!]
\flushleft{\large (a)}\\
\includegraphics[width=0.37\textwidth,clip]{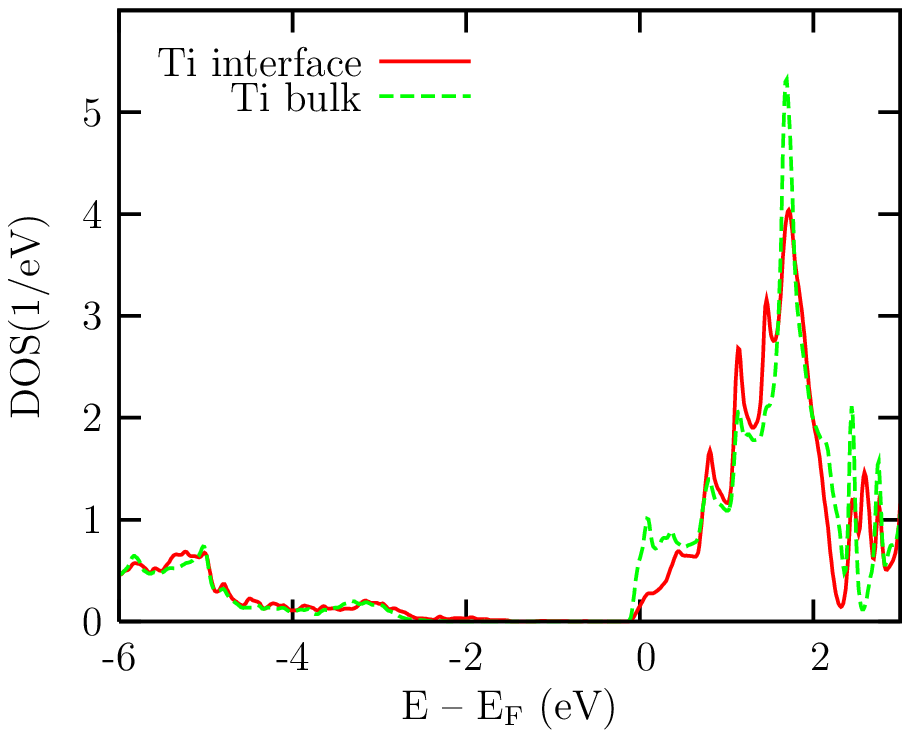}
\vspace{-0.5cm}
\flushleft{\large (b)}\\
\includegraphics[width=0.37\textwidth,clip]{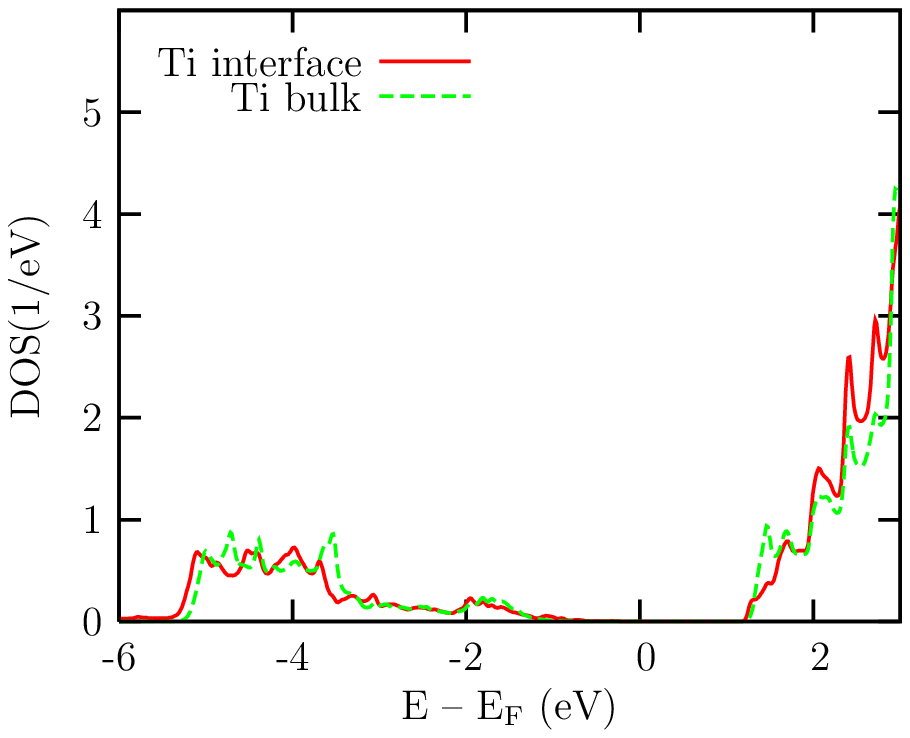}
\vspace{-0.5cm}
\flushleft{\large (c)}\\
\includegraphics[width=0.37\textwidth,clip]{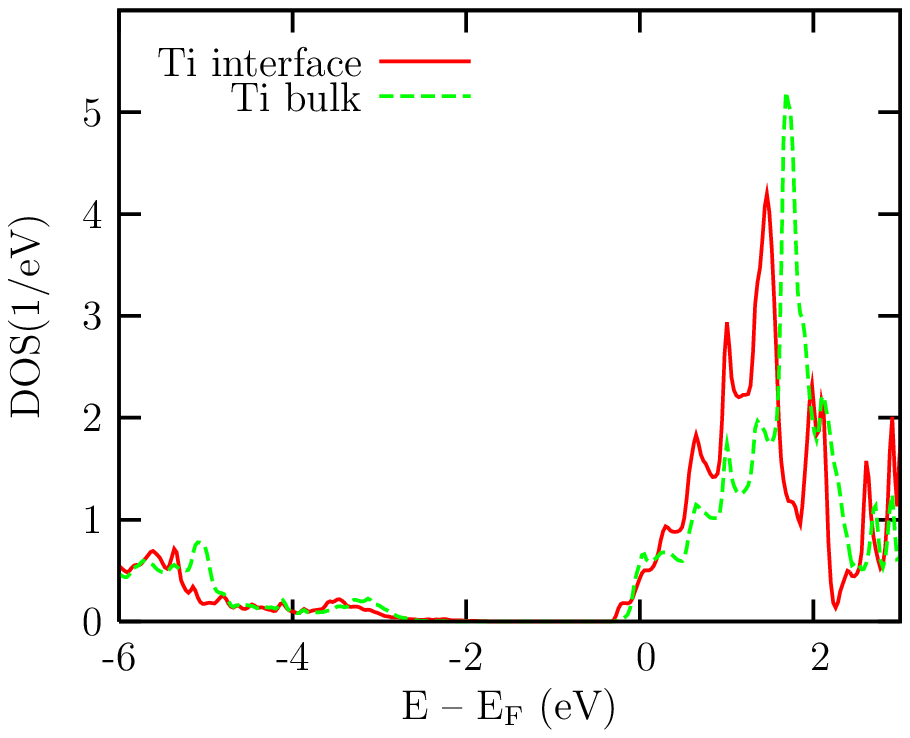}
\caption{Interface Ti $3d$ DOS for (a) the two layer system, accounting only for the
interface structure relaxation, and (b)/(c) the two/four layer system
after full structure relaxation. In each case the results are compared to a
bulk-like Ti atom in the supercell.}
\label{fig_2l}
\end{figure}

We next investigate in detail the band bending in our heterostructure. Since the energy
difference E${\rm_V}$-E${\rm_C}$ at the interface is determined by the two
component materials, it should not vary when the film thickness changes. By
decreasing the vacuum-interface distance, the up-bending at the surface counteracts
the down-bending at the LaAlO$_3$ side of the interface. This reduced down-bending
continues into the SrTiO$_3$ domain. Consequently, the electronic properties of the
interface may be tailored by altering the film thickness. In particular, our
results confirm the observation that a growing thickness of the LaAlO$_3$ surface
layer enhances the metallicity of the heterointerface \cite{thiel06,yoshimatsu08}.
For an increasing film thickness the charge depletion declines
due to electronic screening. As concerns the qualitative dependence on the
thickness of the LaAlO$_3$ layer, theory and experiment reveal an excellent agreement,
while a critical number of three unit cells for reaching an insulator cannot be
reproduced. Instead, an insulating state appears not until the thickness of the
surface layer decreases to two unit cells, see Fig.\ \ref{fig_2l}a.

\begin{figure*}[b]
\includegraphics[width=0.9\textwidth,clip]{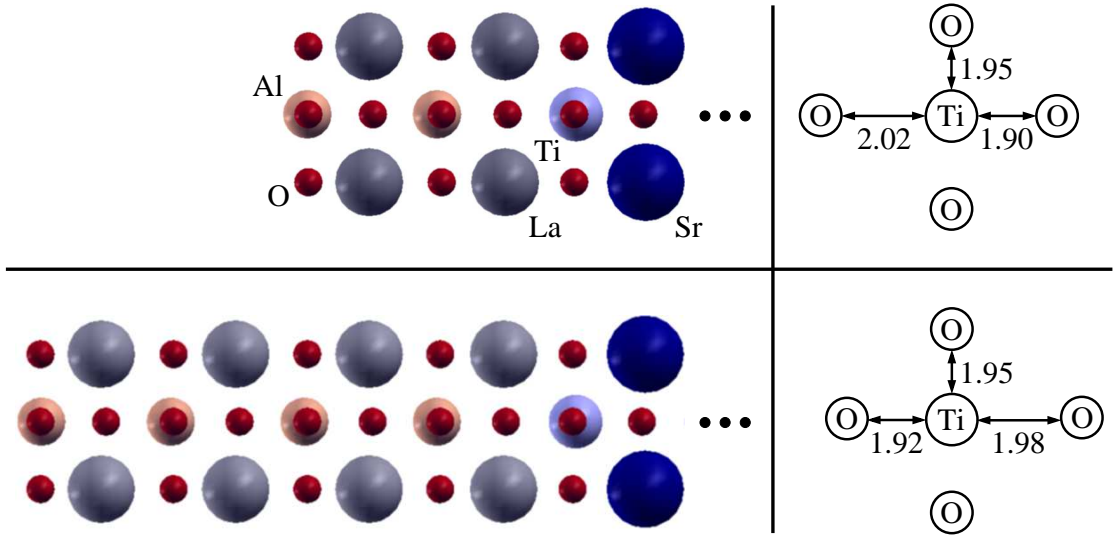}
\caption{Structure of the two (top) and four (bottom) layer systems. The bond lengths in
the interface TiO$_6$ octahedron are illustrated on the right hand side.}
\label{struct-LaAl2}
\end{figure*}

Structural relaxations appear at the LaAlO$_3$/SrTiO$_3$ interface and
in the LaAlO$_3$ layer itself, influencing each other, where drastic
alterations of these patterns could come along with a variation of layer thicknesses.
In this context, we investigate the electronic properties of fully relaxed
thin-film LaAlO$_3$/SrTiO$_3$ heterostructures, comparing configurations
with two and four LaAlO$_3$ surface layers, see Fig.\ \ref{struct-LaAl2}. We
first discuss the two layer system and then turn to the four layer system.

A distinct buckling of the AlO$_2$ plane right at the interface is evident. Furthermore,
according to Fig.\ \ref{struct-LaAl2} (top right) and the bond lengths summarized
in Table \ref{MO-distances}, the TiO$_6$ octahedra are distorted. The
in-plane M-O distances are nearly constant with values of 1.95\,\AA\ (M=Ti,Al) and
2.76\,\AA\ (M=Sr,La). For the ultra-thin two layer system the distortion pattern of
the TiO$_6$ octahedra is reversed to that observed in the multi-layer system: The
elongated octahedral Ti-O bond is not oriented to the substrate but to the film. 
The back-draw of this type of distortion on the electronic states is addressed
in Fig.\ \ref{fig_2l}b,c. When we take into account only the interface structure
relaxation of the multi-layer system a small but finite density of states (DOS) remains
at the Fermi level in Fig.\ \ref{fig_2l}a. In contrast, a full relaxation of the atomic
positions results in a clearly insulating state in Fig.\ \ref{fig_2l}b.
On the other hand, we find that the four layer system shows a rather smooth alignment
of the AlO$_2$ planes. The deformation pattern of the TiO$_6$ octahedra is again
characterised by one elongated octahedral Ti-O bond, which now is oriented towards the
substrate, as in the multi-layer system, compare Fig.\ \ref{struct-LaAl2} (bottom right).
This type of distortion results in metallicity as depicted in Fig.\ \ref{fig_2l}c.

\begin{table*}[t]
\begin{tabular}{l|c|c|c|c|c|c}
&layers&Ti-O, LaAlO$_3$ side&Ti-O, SrTiO$_3$ side& La-O, interface & La-O, surface & Al-O, surface\\\hline
start & & 1.97 & 1.94 & 2.67, 2.79 & 2.62, 2.86 & 1.88\\
end & two & 1.92 & 1.98 &2.53, 2.87&  2.49, 3.00 &1.85 \\
end & four & 2.02 & 1.90 & 2.74, 2.77 & 2.68, 2.82 & 1.90\\
\end{tabular}
\caption{\label{MO-distances} Start and end values of the structure optimization
(selected bond lengths).}
\end{table*}

In summary, we find a strong structure relaxation for thin-film
LaAlO$_3$/SrTiO$_3$ heterostructures. At the contact of the component
materials, the anions and cations are displaced in opposite directions,
leading to a characteristic distortion pattern of the TiO$_6$
octahedra. This distortion increases the carrier density within the
metallic inter-layer. Variation of the thickness of the LaAlO$_3$
surface slab, while keeping the interface structure fixed, shows that
conduction states are subject to a rigid band shift due to a directed
charge transfer. The Fermi level is modified at the heterointerface,
since surface electronic states leak out into the vacuum. Particularly,
the charge transfer towards the surface counteracts the intrinsic
interface doping. However, the structural changes are subtle, both
at the interface and in the LaAlO$_3$ layer itself.

The interplay between surface and interface relaxation becomes efficient when
the full substrate-film system is optimized. Lowering of the LaAlO$_3$ film
thickness below four unit cells then leads to an inversion of the TiO$_6$
distortion pattern at the heterointerface. This new pattern leads to an
insulating instead of a metallic state. Therefore, the transition from
insulating to conducting behaviour seems to be triggered by both the strong
dependence of the interface charge density on the film thickness and the sharp
turn-over of the associated interface lattice distortion. Lee and MacDonald
\cite{lee07} report on a minimum number of polar layers necessary for inducing
an electronic reconstruction, which could likewise explain a critical
thickness. However, the strong response of the interface crystal structure
to the film thickness rather points to a structural mechanism.

In conclusion, taking into account the full lattice relaxation, band structure
calculations are able to reproduce the experimental critical film thickness for a
conducting LaAlO$_3$/SrTiO$_3$ interface even {\it quantitatively}. It is to be
expected that the combined mechanism described in this work depends less on the
atomic species at the interface than on the specific structural conditions.
Consequently, it is quite general and might be applied to various oxide
heterointerfaces, paving the way for a systematic design of their electronic
properties in thin-film configuration.

\subsection*{Acknowledgement}
We acknowledge fruitful discussions with P.\ Schwab.
The work was supported by the Deutsche Forschungsgemeinschaft (SFB 484).

\end{document}